\documentclass[12pt]{article}
\def\nabstar#1{\nabla\kern-0.5pt\smash{\raise 4.5pt\hbox{$\ast$}}
               \kern-4.5pt_{#1}}

\def\drvstar#1{\partial\kern-0.5pt\smash{\raise 4.5pt\hbox{$\ast$}}
               \kern-5.0pt_{#1}}

\def\newline{\relax\ifhmode\null\hfil\break\else\nonhmodeerr@\newline\fi}
\def\frac#1#2{{#1\over#2}}
\def\text#1{{\hbox{\rm #1}}}
\def\flushpar{{\par \noindent}}

\newcommand{\beq}{\begin{equation}}
\newcommand{\eeq}{\end{equation}}
\newcommand{\bea}{\begin{eqnarray}}
\newcommand{\eea}{\end{eqnarray}}
\def\Id{ \mbox{1\hspace{-1.2mm}I} }
\def\EQ{\hspace{-2mm} &=& \hspace{-2mm}}

\def\BA{\begin{eqnarray}}
\def\EA{\end{eqnarray}}
\def\BAN{\begin{eqnarray*}}
\def\EAN{\end{eqnarray*}}

\def\nn{\nonumber\\}

\def\tr{\mbox{tr}}

\def\det{\mbox{det}}

\def\gm5{\gamma_5}

\input epsf.sty
\newdimen\psfigsize
\def\psfigure#1 #2 #3 #4 #5{
    \begin{figure}[tbh]
      \begin{center}
      \vbox{
        \null\vskip-0.2in\hskip#2
        \epsfxsize=#1
        \epsfbox{#4}
        \vskip -0.3in
        \caption {#5 \label{#3}}
        \vskip 0.0 true in plus 0.3 true in
      }
      \end{center}
   \end{figure}
}
\usepackage{graphicx}
\begin{document}
\thispagestyle{empty}
\begin{flushright}
NTUTH-05-505C \\
June 2005 \\
\end{flushright}
\bigskip\bigskip\bigskip
\vskip 2.5truecm
\begin{center}
{\LARGE {Pseudoscalar Decay Constants 
$ f_{D} $ and $ f_{D_s} $ in Lattice QCD  
with Exact Chiral Symmetry }}
\end{center}
\vskip 1.0truecm
\begin{center}
{Ting-Wai Chiu, Tung-Han Hsieh, Jon-Yu Lee, Pei-Hua Liu, Hsiu-Ju Chang}
\end{center}
\vskip5mm
\centerline{Department of Physics and}
\centerline{National Center for Theoretical Sciences at Taipei,}
\centerline{National Taiwan University, Taipei 106, Taiwan}
\vskip 1cm
\bigskip \nopagebreak \begin{abstract}
\noindent

We determine the masses and decay constants of pseudoscalar mesons 
$ D $, $ D_s $, and $ K $  
in quenched lattice QCD with exact chiral symmetry.
For 100 gauge configurations generated with single-plaquette 
action at $ \beta = 6.1 $ on the $ 20^3 \times 40 $ lattice,
we compute point-to-point quark propagators for 30 quark masses
in the range $ 0.03 \le m_q a \le 0.80 $, and measure the
time-correlation functions of pseudoscalar and vector mesons.
The inverse lattice spacing $ a^{-1} $
is determined with the experimental input of $ f_\pi $,
while the strange quark bare mass $ m_s a = 0.08 $, and the charm
quark bare mass $ m_c a = 0.80 $ are fixed such that the masses of
the corresponding vector mesons are in good agreement with  
$ \phi(1020) $ and $ J/\psi(3097) $ respectively.
Our results of pseudoscalar-meson decay constants
are $ f_K = 152(6)(10) $ MeV, $ f_D = 235(8)(14)$ MeV, 
and $ f_{D_s} = 266(10)(18) $ MeV.



\end{abstract}
\vskip 1.5cm 
\newpage\setcounter{page}1

\section{Introduction}

The pseudoscalar-meson decay constants 
(e.g., $ f_D $, $ f_{D_s} $, $ f_B $ and $ f_{B_s} $)  
play an important role in extracting the CKM matrix elements   
(e.g., the leptonic decay width of $ D_s^+ \to l^+ \nu_l $ is 
proportional to $ f_{D_s}^2 | V_{cs} |^2 $),  
which are crucial for testing the flavor sector of the 
standard model via the unitarity of CKM matrix.
Experimentally\footnote{See, for example, Refs. 
\cite{Bonvicini:2004gv,Ablikim:2004ry,Bai:1994qz,Kodama:1996xq}, 
and other experimental results complied by PDG \cite{Eidelman:2004wy}.}, 
precise determination of $ f_{D_s} $ 
and $ f_D $ will soon result from the high-statistics program 
of CLEO-c, however, the determination of $ f_{B} $ and $ f_{B_s} $
remains beyond the reach of current experiments. 

Theoretically, lattice QCD provides a solid framework to compute 
the masses and decay constants of pseudoscalar mesons 
(as well as other physical observables) nonperturbatively 
from the first principles of QCD. 
Thus reliable lattice QCD determinations of $ f_B $ and $ f_{B_s} $ 
are of fundamental importance, in view of their experimental 
determinations are still lacking. Obviously, the first step  
for lattice QCD is to check whether the lattice determinations of 
$ f_D $ and $ f_{D_s} $ will agree with those coming soon from 
the high-statistics charm program of CLEO-c. This motivates 
our present study.          

In this paper, we compute quenched quark propagators 
for 30 quark masses in the range $ 0.03 \le m_q a \le 0.80 $,      
in the framework of optimal domain-wall fermion  
proposed by Chiu \cite{Chiu:2002ir}-\cite{Chiu:2003ir}. 
Then we determine the inverse lattice spacing $ a^{-1} = 2.237(76) $ GeV 
from the pion time-correlation function, with the experimental input 
of pion decay constant $ f_\pi = 132 $ MeV. 
The strange quark bare mass $ m_s a = 0.08 $ 
and the charm quark bare mass $ m_c a = 0.80 $ are fixed  
such that the corresponding masses  
extracted from the vector meson correlation 
function agree with   
$ \phi(1020) $ and $ J/\psi (3097) $ respectively. 
Then the masses and decay constants of any hadrons
containing $ c, s $, and $ u (d) $ quarks\footnote{In this paper, we work in 
the isospin limit $ m_u = m_d $.} 
are predictions of QCD from the first principles,  
with the understanding that 
chiral extrapolation to physical $ m_{u,d} \simeq m_s/25 $
(or equivalently $ m_\pi = 135 $ MeV) is required  
for any observables containing $ u(d) $ quarks. 

For pseudoscalar and vector mesons, we measure their 
time correlation functions for the following three categories:
(i) two quarks have the same mass;
(ii) one quark mass is fixed at $ m_s $;    
(iii) one quark mass is fixed at $ m_c $.    
Note that for mesons which are composed of strange and/or charm quarks, 
their masses and decay constants can be measured directly 
without chiral extrapolation. 

The outline of this paper is as follows. In section 2, we outline 
our formulation of exact chiral symmetry on the lattice, and our 
computation of quark propagators. In section 3, we determine
the inverse lattice spacing spacing, the strange quark bare mass, 
and the charm quark bare mass. In section 4, we present our
results of $ m_K $ and $ f_K $. In section 5, we present
our results of $ m_D $, $ m_{D_s} $, $ f_D $, and $ f_{D_s} $. 
In section 6, we summarize our results and conclude with some remarks.  

\section{Lattice quarks with exact chiral symmetry}

To implement exact chiral symmetry on the lattice 
\cite{Kaplan:1992bt,Narayanan:1995gw,Neuberger:1997fp,Ginsparg:1981bj}, 
we consider the optimal domain-wall fermion proposed by 
Chiu \cite{Chiu:2002ir}-\cite{Chiu:2003ir}. 
The action of optimal domain-wall fermion can be written as \cite{Chiu:2003ir}
\bea
\label{eq:ODWF}
{\cal A}_F \EQ 
\sum_{s,s'=0}^{N_s+1} \sum_{x,x'}
\bar\psi(x,s)
\{ (\omega_s D_w(x,x') + \delta_{x,x'}) \delta_{ss'} \nn
&& \hspace{12mm} + (\omega_s D_w(x,x') - \delta_{x,x'})
    (P_{+} \delta_{s',s-1} + P_{-} \delta_{s',s+1} ) \} \psi_(x',s')
\eea
with boundary conditions
\BAN
\label{eq:bc1a}
P_{+} \psi(x,-1) \EQ -r \ m_q \ P_{+} \psi(x,N_s+1), \\
\label{eq:bc2a}
P_{-} \psi(x,N_s+2) \EQ -r \ m_q \ P_{-} \psi(x,0), 
\hspace{4mm} r=\frac{1}{2m_0}, 
\EAN
where $ m_q $ is the bare quark mass,  
and $ \{ \omega_s, s = 0, \cdots, N_s+1 \} $  
are specified by the exact formula derived in Ref. \cite{Chiu:2002ir}. 
Here $ H_w = \gamma_5 D_w $, and $ D_w $ is the standard Wilson Dirac 
operator plus a negative parameter $ -m_0 $ ($ 0 < m_0 < 2 $). 
The quark fields are
constructed from the boundary modes at $ s=0 $ and $ s=N_s + 1 $
with $ \omega_0 = \omega_{N_s + 1} = 0 $ \cite{Chiu:2003ir}:
\bea
\label{eq:q}
q(x) &=& \sqrt{r} \left[ P_{-} \psi(x,0) + P_{+} \psi(x,N_s+1) \right], \\ 
\label{eq:qbar}
\bar q(x) &=& \sqrt{r}
\left[ \bar\psi(x,0) P_{+} + \bar\psi(x,N_s+1) P_{-} \right].
\eea
After introducing pseudofermions with $ m_q = 2 m_0 $, 
the generating functional for $n$-point Green's function
of the quark fields can be derived as \cite{Chiu:2003ir}, 
\bea
\label{eq:ZW_odwf}
Z[J,\bar J] =
\frac{\int [dU] e^{-{\cal A}_g} \det [(D_c+m_q)(1+rD_c)^{-1}]
           \exp \left\{ \bar J (D_c+ m_q)^{-1} J \right\}  }
     {\int [dU] e^{-{\cal A}_g} \det [(D_c + m_q)(1+r D_c)^{-1}] }
\eea
where $ {\cal A}_g $ is the action of the gauge fields,
$ \bar J $ and $ J $
are the Grassman sources of $ q $ and $ \bar q $ respectively, and
\bea
\label{eq:Dc}
D_c \EQ 2 m_0 \frac{ 1 + \gamma_5 S_{opt} }{ 1 - \gamma_5 S_{opt} } \ , \\
S_{opt} \EQ \frac{1 - \prod_{s=1}^{N_s} T_s}
                {1 + \prod_{s=1}^{N_s} T_s}, \\
 T_s \EQ \frac{1 - \omega_s H_w }{1 + \omega_s H_w} . 
\eea
Using the exact formula of $ \omega_s $ \cite{Chiu:2002ir}, 
one immediately obtains 
\bea
S_{opt} = \left\{ \begin{array}{ll}
           H_w R_Z^{(n,n)}(H_w^2),   &  N_s = 2n+1,   \\
           H_w R_Z^{(n-1,n)}(H_w^2), &  N_s = 2n,    \\
                  \end{array} \right.
\label{eq:S_opt_RZ}
\eea
where $ R_Z(H_w^2) $ is the Zolotarev optimal rational polynomial
\cite{Akhiezer:1992} for the inverse square root of $ H_w^2 $,  
\bea
\label{eq:rz_nn}
R^{(n,n)}_Z(H_w^2) &=& \frac{d_0}{\lambda_{min}}
\prod_{l=1}^{n} \frac{ 1+ h_w^2/c_{2l} }{ 1+ h_w^2/c_{2l-1} } \nn
&=& \frac{1}{\lambda_{min}} (h_w^2 + c_{2n})
    \sum_{l=1}^n \frac{b_l}{h_w^2 + c_{2l-1}} \ ,
\hspace{8mm}  h_w^2 = H_w^2/\lambda_{min}^2
\eea
and
\bea
\label{eq:rz_n1n}
R^{(n-1,n)}_Z(H_w^2) = \frac{d'_0}{\lambda_{min}}
\frac{ \prod_{l=1}^{n-1} ( 1+ h_w^2/c'_{2l} ) }
     { \prod_{l=1}^{n} ( 1+ h_w^2/c'_{2l-1} ) }
= \frac{1}{\lambda_{min}} \sum_{l=1}^n \frac{b'_l}{h_w^2 + c'_{2l-1}} \ ,
\eea
where the coefficients $ d_0 $, $ d'_0 $, $ c_l $ and $ c'_l $
are expressed in terms of elliptic functions \cite{Akhiezer:1992}
with arguments depending on $ N_s $
and $ \lambda_{max}^2 / \lambda_{min}^2 $,  
and $ \lambda_{min} $ ($ \lambda_{max} $) is fixed   
to be the greatest lower bound (least upper bound) of the
eigenvalues of $ |H_w| $ for the set of gauge configurations
under investigation.

From (\ref{eq:ZW_odwf}), the effective 4D lattice Dirac operator
for the fermion determinant is
\bea
\label{eq:Dm_odwf}
D(m_q) = (D_c+m_q)(1+rD_c)^{-1}
       = m_q + (m_0 - m_q/2) \left[ 1 + \gamma_5 H_w R_Z(H_w^2) \right]
\eea
and the quark propagator in background gauge field is
\bea
\label{eq:quark_prop}
\langle q(x) \bar q(y) \rangle \EQ
- \left. \frac{\delta^2 Z[J,\bar J]}{\delta \bar J(x) \delta J(y)}
\right|_{J=\bar J=0}                                             \nn
\EQ (D_c + m_q)^{-1}_{x,y}=(1-rm_q)^{-1}[D^{-1}_{x,y}(m_q) - r \delta_{x,y}]
\eea
Note that $ D_c $ is exactly chirally symmetric 
(i.e. $ D_c \gamma_5 + \gamma_5 D_c = 0 $) in the limit $ N_s \to \infty $, and its deviation from exact chiral 
symmetry due to finite $ N_s $ is the {\it minimal} provided that  
the weights $ \{ \omega_s \} $ are fixed according to the formula 
derived in Ref. \cite{Chiu:2002ir}. Further, 
the bare quark mass $ m_q $ (whether heavy or light) 
in the quark propagator $ (D_c + m_q)^{-1} $ 
is well-defined for any gauge configurations.   

In practice, we have two ways to evaluate the quark propagator 
(\ref{eq:quark_prop}) in background gauge field: \\
(i) To solve the linear system of the 5D optimal
DWF operator; \\
(ii) To solve $ D^{-1}_{x,y}(m_q) $ from the system
\bea
\label{eq:Dm_RZ}
D(m_q) Y =
\left[m_q+(m_0-m_q/2) \left(1+\gamma_5 H_w R_Z(H_w^2)\right) \right] Y
= \Id \ ,
\eea
with nested conjugate gradient \cite{Neuberger:1998my}, and then
substitute the solution vector $ Y $ into (\ref{eq:quark_prop}).

Since either (i) or (ii) yields exactly the same quark propagator,
in principle, it does {\it not} matter which linear system one actually
solves. However, in practice, one should choose the most efficient scheme
for one's computational system (hardware and software).
For our present system (a Linux PC cluster of 100 nodes \cite{Chiu:2002bi}),  
it has been geared to the scheme (ii), and it attains the maximum efficiency
if the inner conjugate gradient loop of (\ref{eq:Dm_RZ}) is iterated with 
Neuberger's 2-pass algorithm \cite{Neuberger:1998jk}.
So we use the scheme (ii) to compute the quark propagator,
with the quark fields (\ref{eq:q})-(\ref{eq:qbar})
defined by the boundary modes at $ s=0 $ and $ s=N_s + 1 $. 
Note that Neuberger's 2-pass algorithm 
not only provides very high precision of chiral symmetry with fixed 
amount of memory, but also is faster than the single pass algorithm 
for $ n > 12 \sim 25 $ (where $ n $ is the order of the rational polynomial 
$ R^{(n-1,n)} $) for most computer platforms, as discussed by 
Chiu and Hsieh \cite{Chiu:2003ub}.

We generate 100 gauge configurations with single plaquette gauge action
at $ \beta = 6.1 $ on the $ 20^3 \times 40 $ lattice.
Fixing $ m_0 = 1.3 $, we project out 16 low-lying eigenmodes of 
$ |H_w| $ and perform the nested conjugate gradient in the complement
of the vector space spanned by these eigenmodes. For  
$ N_s = 128 $, 
the weights $ \{ \omega_s \} $ are fixed with $ \lambda_{min} = 0.18 $ 
and $ \lambda_{max} = 6.3 $, 
where $ \lambda_{min} \le \lambda(|H_w|) \le \lambda_{max} $
for all gauge configurations.    
For each configuration, point to point quark propagators are computed
for 30 bare quark masses in the range $ 0.03 \le m_q a \le 0.8 $, 
with stopping criteria $ 10^{-11} $ and $ 2 \times 10^{-12} $
for the outer and inner conjugate gradient loops respectively.
Then the norm of the residual vector of each column of the quark propagator 
is less than $ 2 \times 10^{-11} $  
\BAN
|| (D_c + m_q ) Y - \Id || < 2 \times 10^{-11},  
\EAN
and the chiral symmetry breaking due to finite $ N_s $ is 
less than $ 10^{-14} $,
\BAN
\sigma = \left| \frac{Y^{\dagger} S^2 Y}{Y^{\dagger} Y} - 1 \right|
< 10^{-14},
\EAN
for every iteration of the nested conjugate gradient. 
Further details of our scheme have been described in 
Refs. \cite{Chiu:2002xm,Chiu:2003ub}.

In this paper, we measure the time-correlation functions 
for pseudoscalar ($PS$) and vector ($V$) mesons, 
\bea
\label{eq:CPS}
C_{PS} (t) \EQ  
\left< 
\sum_{\vec{x}}
\tr\{ \gamma_5 (D_c + m_Q)^{-1}_{x,0} \gamma_5 (D_c + m_q)^{-1}_{0,x} \} 
\right>_U   \\
\label{eq:CV}
C_V (t) \EQ \left< 
\frac{1}{3} \sum_{\mu=1}^3 \sum_{\vec{x}}
\tr\{ \gamma_\mu (D_c + m_Q)^{-1}_{x,0} \gamma_\mu
     (D_c + m_q)^{-1}_{0,x} \} \right>_U
\eea
where the subscript $ U $ denotes averaging over gauge configurations. 
Here $ C_{PS}(t) $ and $ C_V(t) $ are measured for the following 
three categories: \\
(i) Symmetric masses $ m_Q = m_q $ \ ,  \\
(ii) Asymmetric masses with fixed $ m_Q = m_s = 0.08 a^{-1} $ \ , \\    
(iii) Asymmetric masses with fixed $ m_Q = m_c = 0.80 a^{-1} $ \ , \\
where $ m_q $ is varied for 30 masses in the range 
$ 0.03 \le m_q a \le 0.80 $.

\section{Determination of $ a^{-1} $, $ m_s $, and $ m_c $}

\begin{figure}[htb]
\begin{center}
\hspace{0.0cm}\includegraphics*[height=12cm,width=10cm]{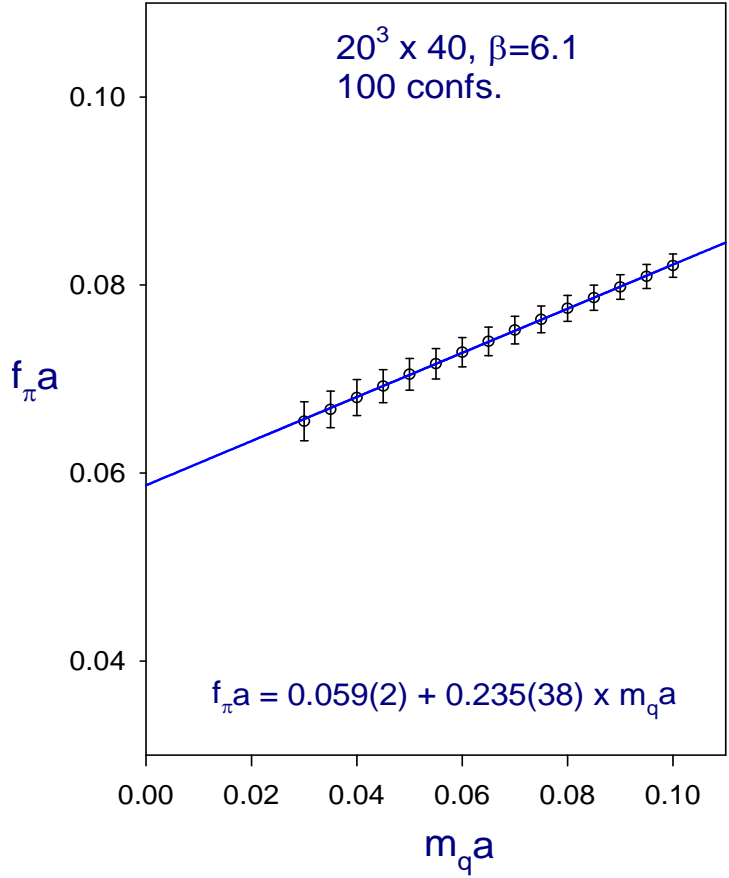}
\caption{The pion decay constant $ f_\pi a $ versus the
bare quark mass $ m_q a $. The solid line is the linear fit.}
\label{fig:fpi_20}
\end{center}
\end{figure}

\begin{figure}[htb]
\begin{center}
\hspace{0.0cm}\includegraphics*[height=12cm,width=10cm]{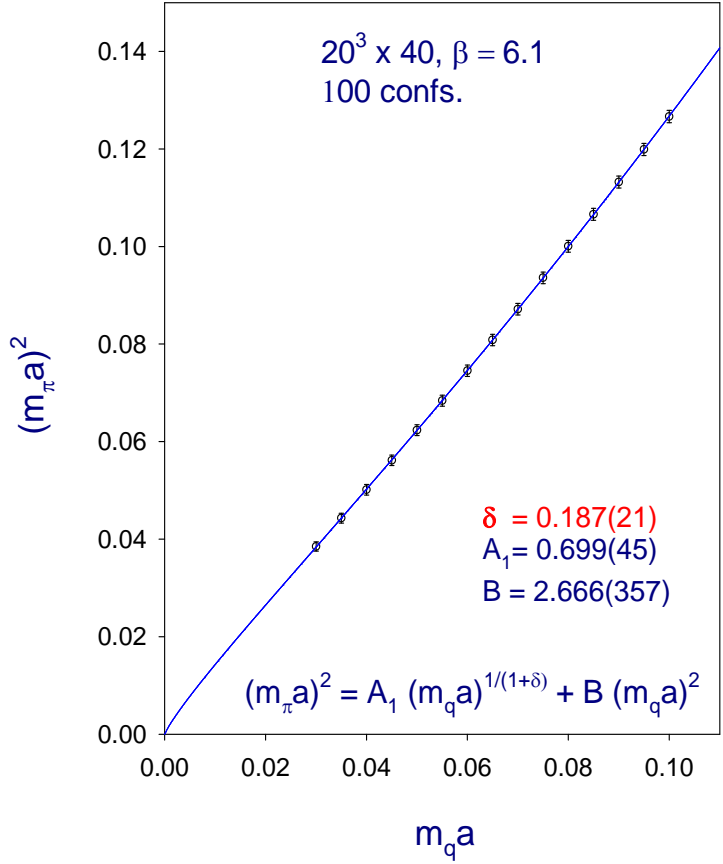}
\caption{The pion mass square $ (m_\pi a)^2 $ versus the
bare quark mass $ m_q a $. The solid line is the fit of Eq. (\ref{eq:mpi2_c1}).}
\label{fig:mpi2_20}
\end{center}
\end{figure}

For symmetric masses $ m_Q = m_q $, the pseudoscalar 
time-correlation function $ C_\pi (t) $ (\ref{eq:CPS}) is measured, 
and is fitted to the usual formula 
\bea
\label{eq:Gt_fit}
\frac{Z}{2 m_{\pi} a } [ e^{-m_{\pi} a t} + e^{-m_{\pi} a (T-t)} ]
\eea
to extract the pion mass $ m_{\pi} a $ and the pion decay constant
\bea
\label{eq:fpi}
f_{\pi} a = 2 m_q a \frac{\sqrt{Z}}{m_{\pi}^2 a^2 } \ .
\eea

In Figs. \ref{fig:fpi_20} and \ref{fig:mpi2_20},
we plot the decay constant $ f_\pi a $ and pion mass square $ (m_\pi a)^2 $  
versus bare quark mass $ m_q a $, respectively.

The data of $ f_{\pi} a $ (see Fig. \ref{fig:fpi_20}) is well fitted by
the straight line
\BAN
f_{\pi} a = 0.059(2) + 0.235(38)  \times  (m_q a) \ . 
\EAN
Then taking $ f_\pi a $ at $ m_q a = 0 $
equal to $ 0.132 $ GeV times the lattice spacing $ a $,
we can determine the lattice spacing $ a $ and its inverse,
\bea
a^{-1} &=& \frac{0.132}{f_0} \mbox{ GeV} = 2.237(76) \mbox{ GeV} \ , \nn
a &=& 0.088(3) \mbox{ fm} \ .
\label{eq:a}
\eea
Thus the size of our lattice is about 
$ (1.8 \mbox{ fm})^3 \times 3.6 \mbox{ fm} $.
Since the smallest pion mass is $ 439 \mbox{ MeV} $, the lattice
size is about $ (3.9)^3 \times 7.8 $, in units of the Compton wavelength
($\sim0.45 \mbox{ fm}$) of the smallest pion mass. 

The data of $ m_\pi^2 $ (see Fig. \ref{fig:mpi2_20}) can be 
fitted by the form \cite{Sharpe:1992ft}
\bea
\label{eq:mpi2_c1}
m_\pi^2 a^2 = A_1 (m_q a)^{\frac{1}{1+\delta}} + B (m_q a)^2
\eea
in quenched chiral perturbation theory (q$\chi$PT). 
The fitted parameters are 
\bea
\label{eq:delta_pi_c}
\delta &=& 0.187(21) \\
\label{eq:A_c}
A_1 &=& 0.669(45) \\
\label{eq:B_c}
B &=& 2.666(357)
\eea
with $\chi^2 $/d.o.f.=0.54.
Evidently, the coefficient of quenched chiral logarithm
$ \delta = 0.187(21) $ is in good agreement with the theoretical
estimate $ \delta \simeq 0.176 $ in q$\chi$PT.

The bare mass of strange quark is determined by extracting the
mass of vector meson from the time-correlation function
$ C_V(t) $.  At $ m_q a = 0.08 $, $ m_V a = 0.460(4) $,
which gives $ m_V = 1029(10) $ MeV, in good agreement with
the mass of $ \phi(1020) $. Thus we take the strange quark
bare mass to be $ m_s a = 0.08 $.
Similarly, at $ m_q a = 0.80 $, $ m_V a = 1.368(2) $, 
which gives $ m_V = 3060(5) $ MeV, in 
good agreement with the mass of $ J/\Psi(3097) $. 
Thus, we fix the charm quark bare mass to be $ m_c a = 0.80 $.

\section{$f_K$ and $ m_K $}

\begin{figure}[htb]
\begin{center}
\hspace{0.0cm}\includegraphics*[height=12cm,width=10cm]{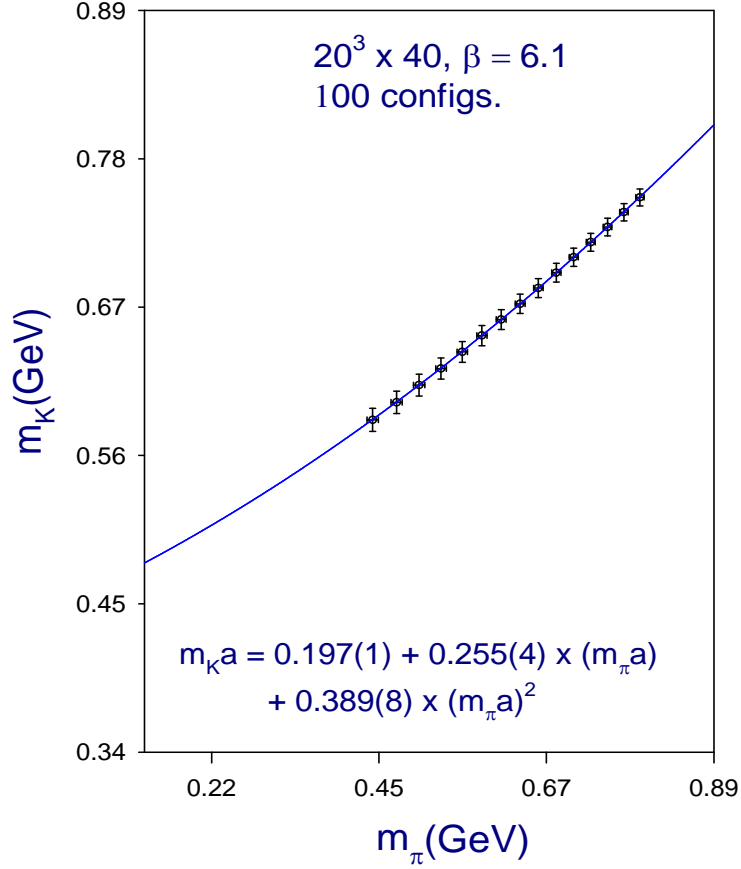}
\caption{The kaon mass $ m_K $ versus
the pion mass $ m_{\pi} $ for 15 bare quark masses 
in the range $ 0.03 \le m_q a \le 0.10 $.
The solid line is the quadratic fit.}
\label{fig:mK_20}
\end{center}
\end{figure}

\begin{figure}[htb]
\begin{center}
\hspace{0.0cm}\includegraphics*[height=12cm,width=10cm]{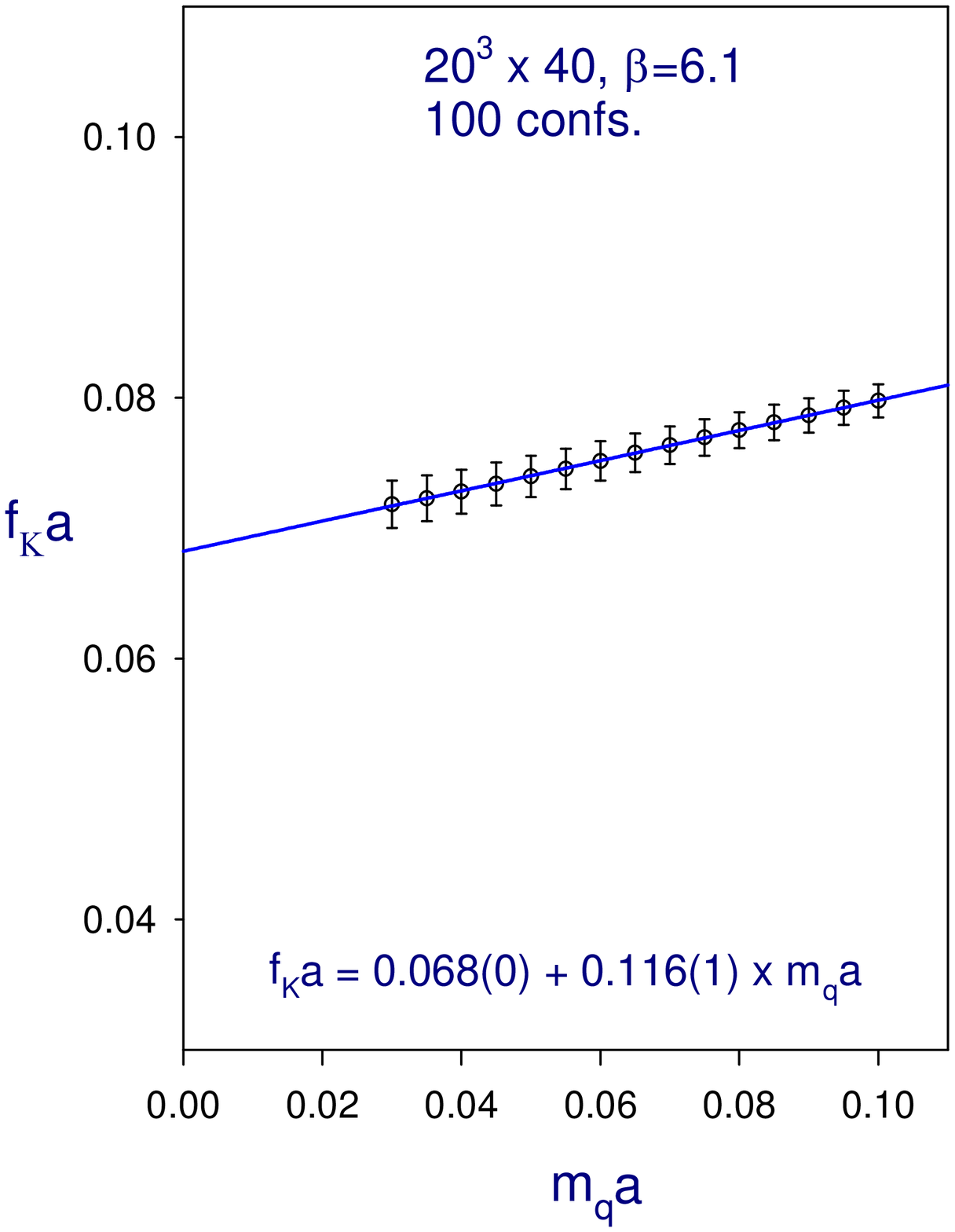}
\caption{The kaon decay constants $ f_K a $  
versus the bare quark mass $ m_q a $. The solid line is the linear fit.}
\label{fig:fK_20}
\end{center}
\end{figure}

Next we measure the time-correlation function of kaon
$ C_{K}(t) $ (\ref{eq:CPS}) with $ m_Q $ fixed at 
$ m_s = 0.08 a^{-1} $, while $ m_q $
is varied for 30 masses in the range $ 0.03 \le m_q a \le 0.80$.
Then the data of $ C_K(t) $ 
is fitted by the formula analogous to (\ref{eq:Gt_fit})
to extract the kaon mass $ m_{K} a $ and the kaon decay constant
$ f_K a $.

In Fig. \ref{fig:mK_20}, the kaon mass $ m_K $ is plotted versus $ m_\pi $, 
for 15 quark masses in the range $ 0.03 \le m_q a \le 0.10 $.  
The data of $ m_K a $ can be fitted by  
\BAN
\label{eq:mK_mpi}
m_K a = 0.197(1) + 0.255(4) (m_\pi a) + 0.389(8) (m_\pi a)^2 \ .
\EAN   
At the physical limit $ m_\pi = 135 $ MeV, it gives 
$ m_K = 478(16) $ MeV, in good agreement with the experimental value 
of kaon mass ($ 495 $ MeV).

In Fig. \ref{fig:fK_20}, $ f_K a $ is plotted versus 
bare quark mass $ m_q a $. 
The data is well fitted by the straight line  
\BAN
f_{K} a = 0.068(0) + 0.116(1)  \times  (m_q a)
\EAN
At $ m_q a = 0 $, it gives $ f_K = 152(6) $ MeV, in agreement
with the value $ f_{K^+} = 159.8 \pm 1.4 \pm 0.44 $ MeV
complied by PDG \cite{Eidelman:2004wy}.

\section{$f_D$, $ f_{D_s} $, $m_D$, and $ m_{D_s} $}

\begin{figure}[htb]
\begin{center}
\hspace{0.0cm}\includegraphics*[height=12cm,width=10cm]{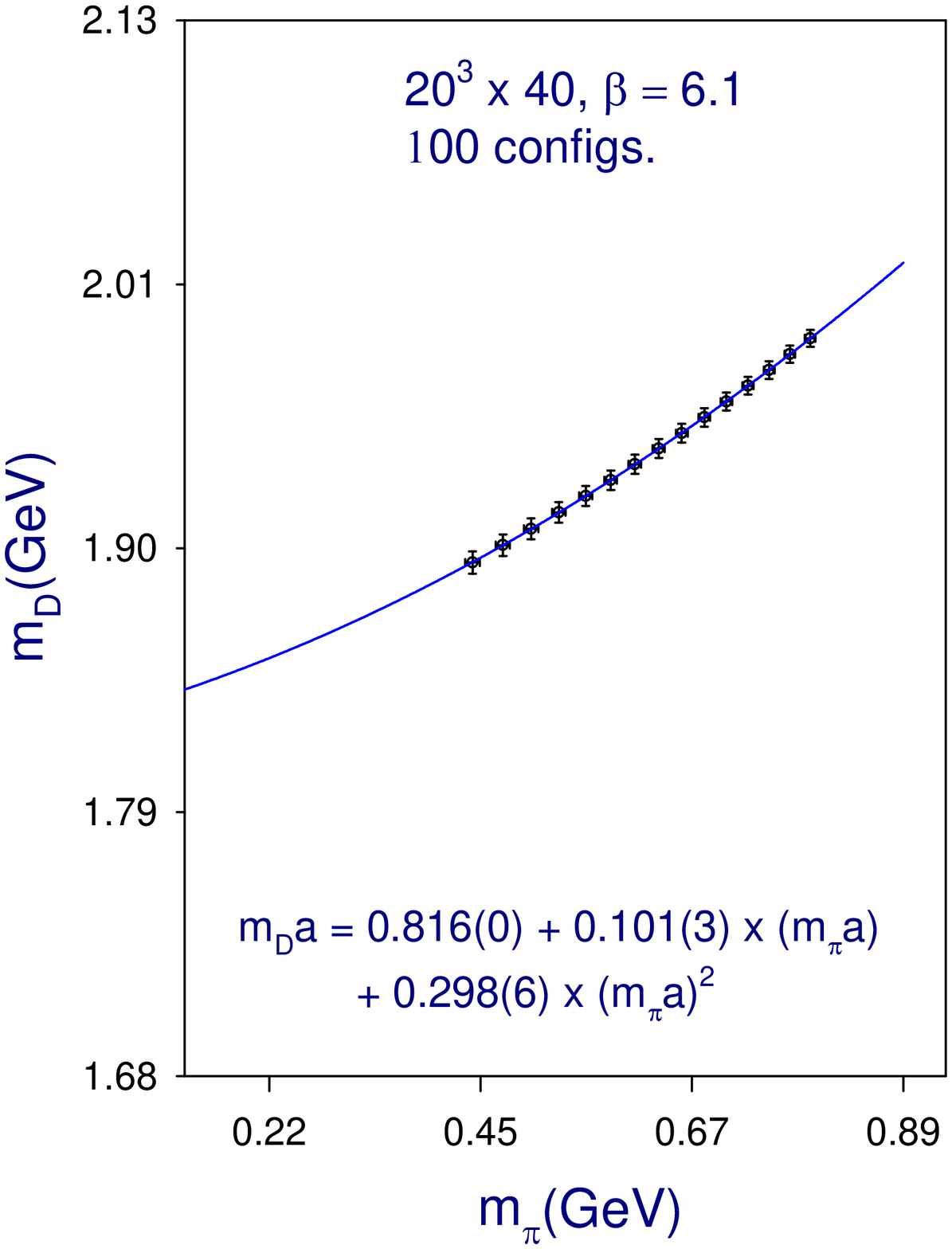}
\caption{The mass of $ D $ meson $ m_D a $ versus the pion mass $ m_{\pi} a $ 
for 15 bare quark masses in the range $ 0.03 \le m_q a \le 0.10 $.
The solid line is the quadratic fit.}
\label{fig:mD_20}
\end{center}
\end{figure}

\begin{figure}[htb]
\begin{center}
\hspace{0.0cm}\includegraphics*[height=12cm,width=10cm]{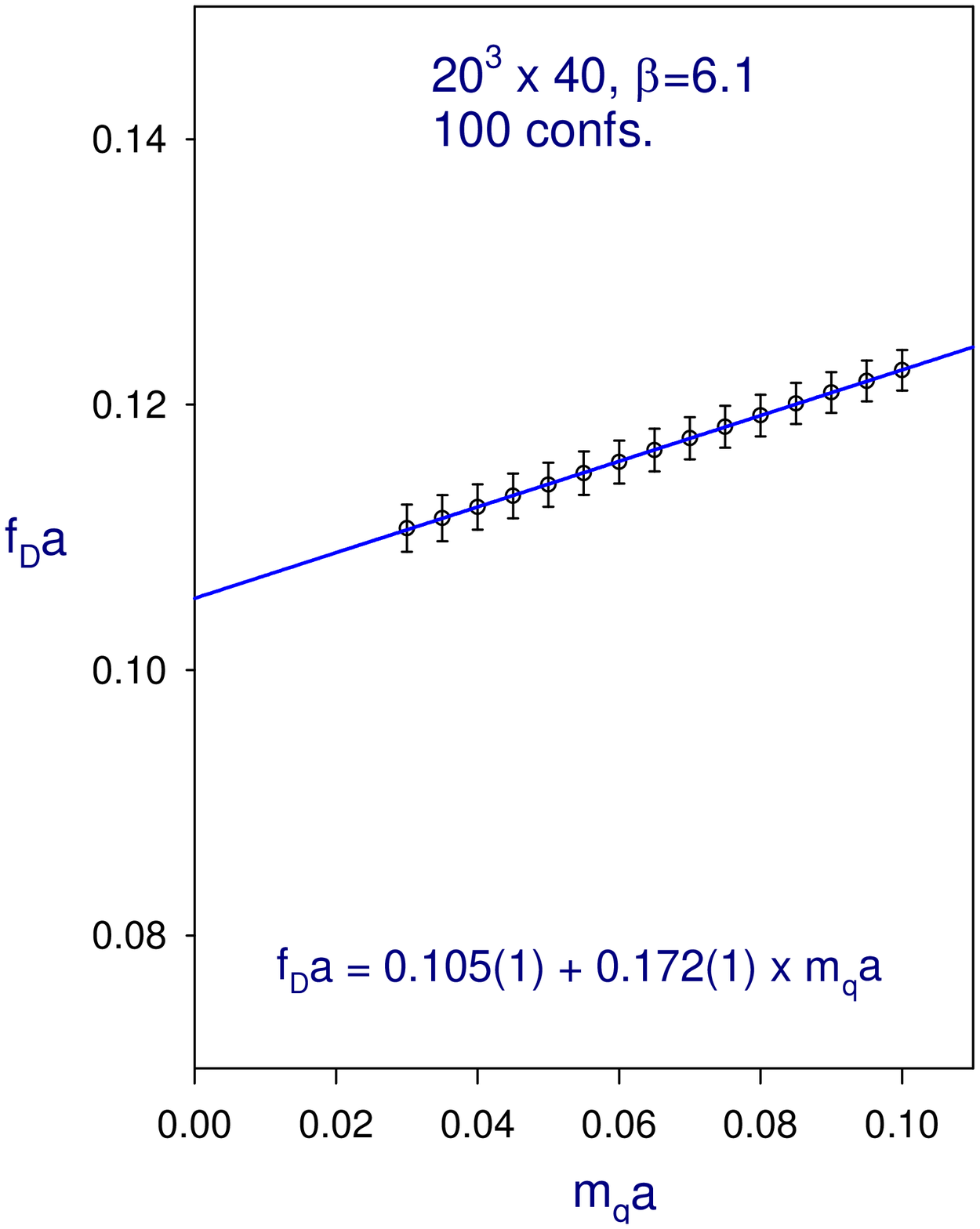}
\caption{The D-meson decay constants $ f_D a $  
versus the bare quark mass $ m_q a $. The solid line is the linear fit.}
\label{fig:fD_20}
\end{center}
\end{figure}

Now we turn to charmed pseudoscalar mesons.  
We measure the time-correlation function  
$ C_{D}(t) $ (\ref{eq:CPS}) with $ m_Q $ fixed at 
$ m_c = 0.80 a^{-1} $, while $ m_q $
is varied for 30 different masses in the range $ 0.03 \le m_q a \le 0.80$.
Then the data of $ C_D(t) $ 
is fitted by the formula analogous to (\ref{eq:Gt_fit})
to extract the mass $ m_{D} a $ and decay constant $ f_D a $. 

In Fig. \ref{fig:mD_20},    
$ m_D a $  is plotted versus $ m_\pi a $, 
for 15 quark masses in the range $ 0.03 \le m_q a \le 0.10 $. 
The data of $ m_D a $ can be fitted by 
\BAN
\label{eq:mD_mpi}
m_D a = 0.816(0) + 0.101(3) (m_\pi a) + 0.298(6) (m_\pi a)^2 
\EAN    
At $ m_\pi = 135 $ MeV, it gives $ m_D = 1842(15) $ MeV, 
in good agreement with the mass of $ D $ meson ($ 1865 $ MeV).
In Fig. \ref{fig:fD_20}, the decay constant $ f_D a $ is plotted 
versus bare quark mass $ m_q a $. 
The data is well fitted by the straight line  
\BAN
f_{D} a = 0.105(1) + 0.172(1)  \times  (m_q a)
\EAN
At $ m_q a = 0 $, it gives $ f_D = 235(8) $ MeV, which serves 
as a prediction of lattice QCD with exact chiral symmetry.

The pseudoscalar meson of $ c \bar s $ or $ s \bar c $ 
corresponds to $ m_Q a = m_c a = 0.80 $ and $ m_q a = m_s a = 0.08 $. 
Its mass and decay constant are extracted directly from the 
time-correlation function, which are plotted as the eleventh 
data point (counting from the smallest one) in Figs. \ref{fig:mD_20}
and \ref{fig:fD_20} respectively.     
The results are $ m_{D_s} a = 0.878(2) $ and $ f_{D_s} a = 0.119(2) $.  
The mass gives $ m_{D_s} = 1964(5) $ MeV, in good agreement with the 
mass of $ D_s(1968) $. The decay constant gives $ f_{D_s} = 266(10) $ MeV,
which agrees with the value $ f_{D_s^+} = 267 \pm 33 $ MeV  
complied by PDG \cite{Eidelman:2004wy}.

\section{Summary and Concluding Remarks}

In this paper, we have determined the masses and decay constants 
of pseudoscalar mesons $ K $, $ D $ and $ D_s $, in quenched lattice QCD
with exact chiral symmetry. Our results are:
\BAN 
m_K \EQ 478 \pm 16 \pm 20 \mbox{ MeV},  \\ 
m_D \EQ 1842 \pm 15 \pm 21 \mbox{ MeV},  \\ 
m_{D_s} \EQ 1964 \pm 5 \pm 10 \mbox{ MeV} ,\\   
f_K \EQ 152 \pm 6 \pm 10 \mbox{ MeV},  \\ 
f_D \EQ 235 \pm 8 \pm 14 \mbox{ MeV},  \\
f_{D_s} \EQ 266 \pm 10 \pm 18 \mbox{ MeV},  
\EAN
where in each case, the first error is statistical, while 
the second is our crude estimate of combined systematic uncertainty.
It is interesting to see whether the values of $ f_D $ and $ f_{D_s} $ 
coming soon from the high-statistics charm program of CLEO-c  
would agree with our values determined by lattice QCD with exact chiral 
symmetry. Further, we note that 
in a recent 3-flavor unquenched lattice QCD study \cite{Simone:2004fr}
with $ O(a^2) $ improved staggered  
light quarks and $ O(a) $-improved charm quark, their  
results of $ f_D $ and $ f_{D_s} $ agree with our values.  

Obviously, our next task is to 
determinate $ f_B $ and $ f_{B_s} $ which are of fundamental importance, 
in view of their experimental determinations are still lacking.  
Since we will not use any approximations for the heavy b quark, 
our lattice spacing must be small enough such that
$ m_b a < 1 $. Even though this does not seem to be formidable for 
$ f_{B_s} $, it is unclear whether we can determine $ f_{B} $ 
reliably via chiral extrapolation. 
Presumably, $ f_{B} $ would behave 
like a function linear in $ m_q $ for a wide range of $ m_q $, similar to 
$ f_D $ (Fig \ref{fig:fD_20}) and $ f_K $ (Fig. \ref{fig:fK_20}), 
then one should be able to obtain a reliable chiral 
extrapolation even for data points with $ m_q > m_s $.

\bigskip
\bigskip

\vfill\eject
\flushpar
{\bf Acknowledgement}

\noindent

\bigskip

This work was supported in part by the National Science Council,
Republic of China, under the Grant No. NSC93-2112-M002-016, and  
by National Center for High Performance Computation at Hsinchu. 
T.W.C. would like to thank Andreas Kronfeld for a timely
remark at the International Conference on QCD and Hadron Physics,
Beijing, June 16-20, 2005.

\bigskip
\bigskip

\vfill\eject

\end{document}